\documentstyle[sprocl,epsf]{article}

\input{psfig}

\def\be{\begin{equation}}
\def\ee{\end{equation}}
\def\beq{\begin{equation}}
\def\eeq{\end{equation}}
\def\bea{\begin{eqnarray}}
\def\eea{\end{eqnarray}}
\def\beqra{\begin{eqnarray}}
\def\eeqra{\end{eqnarray}}

\begin{document}

\title{CRITICAL SLOWING DOWN AND DEFECT FORMATION}

\author{M. PIETRONI}

\address{INFN - Sezione di Padova, via F. Marzolo 8, 
Padova,\\ I-35131, ITALY\\E-mail: pietroni@pd.infn.it} 




\maketitle\abstracts{The formation of topological defects in a second order
phase transition in the early universe is an out-of-equilibrium process. 
Condensed matter experiments seem to support Zurek's mechanism, in
which the freezing of thermal fluctuations close to the critical point
(critical slowing down) plays a crucial role. We discuss how this picture
can be extrapolated to the early universe, pointing out that new scaling
laws may emerge at very high temperatures and showing how critical 
slowing down emerges in the context of a relativistic quantum
field theory.  
}

\section{Zurek's picture of defect formation}
The formation of topological defects (domain walls, strings,...) 
in a second order phase transition is a common phenomenon in condensed 
matter and cosmology. One relevant point in this context is to determine
the initial correlation length of the pattern of defects emerging from the 
critical region or, in other words, to answer the question: 
`when symmetry breaks,
how big are the smallest identifiable pieces?'~\cite{Z}.

Historically, the first answer to this question \cite{K} 
was based on the thermal activation mechanism. In this picture, the 
pattern of defects stabilizes at the Ginzburg temperature, $T_G$, 
below which thermal fluctuations are unable to overcome the free energy 
barrier between inequivalent vacua. 
The initial length scale 
of the pattern of topological defects is then given by the correlation 
length at the Ginzburg temperature, $\xi(T_G)$, which can be estimated 
as
\beq
 \xi(T_G) \sim \frac {1}{\lambda^{1/2} \mu}
\label{ta}
\eeq
where $\mu$ is the mass scale and $\lambda$ the coupling constant 
of the theory. The relevant point about
eq. (\ref{ta}) is its independence on the rate at which temperature is 
changed (quench rate), due to the use of the {\it equilibrium}
free energy from the critical temperature $T_C$ down to $T_G$.

More recently Zurek has proposed a new picture of defect formation \cite{Z}, 
in which dynamical aspects of the phase transition play a key role.
The main ingredient is {\it critical slowing down} (CSD), 
{\it i.e.} the vanishing of the damping  rate of thermal fluctuations 
close to the critical point. CSD can be discussed in
the context of a non-relativistic, classical 
scalar
theory described by an order parameter $\eta(t,\vec{r})$ and a 
Langevin-type equation of motion as
\beq
\frac{\partial \eta(t, \vec{r})}{\partial t} = - \Gamma \frac{\delta {\cal F}}{
\delta \eta(t,\vec{r})} + \zeta(t,  \vec{r}),
\label{LANG}
\eeq
where $ {\cal F}$ is the free-energy,
 $\Gamma$ a phenomenological parameter, and 
$\zeta(t,\vec{r})$ a white noise term. Eq. (\ref{LANG}) 
 has damped plane wave solutions of the form
$\eta (t,\vec{r}) \sim e^{i\,\vec{k}\cdot\vec{r}} \,e^{-\gamma_k t}$,
with 
$\gamma_k = \Gamma [ |\vec{k}|^2 + 2 \alpha 
(T-T_C)]$.
CSD is the statement that
long-wavelength fluctuations ($\vec{k}\to 0$) are not damped as $T \to T_C$, 
which is clearly realized here.

What are the consequences of CSD on the formation of 
topological defects? The scaling $\tau_0 = 1/ \gamma(T) 
\sim (T-T_C)^{-1}$ (where we have defined $ \gamma\equiv  
\gamma_{\vec{k}=0}$)
- obtained from the simple model (\ref{LANG}) - can be generalized to 
\beq
\xi \sim \varepsilon^{-\nu}\,,\,\,\,\,\,\,\,\,\,\,\,\,\,\,
\tau_0 \sim \varepsilon^{-\mu}\,\:\:\:\:\:\:\:{\mathrm for}\,\,\,
 \varepsilon \to 
0\,,
\label{scaling}
\eeq
where $\varepsilon = |(T-T_C)/T_C| = |(t-t_C)/\tau_Q|$ measures the distance
in temperature  - or in time - from the critical point.
As $t_C$ is approached the relaxation time grows until
a certain time $t^*$ at which it becomes larger than the time left before
the phase transition, {\it i.e.}
$\tau_0(t^*) = (t_C - t^*)\,.$
Thermal fluctuations generated from $t^*$ to $t_C$ are then unable to 
relax before 
$t_C$ and the system cannot follow the equilibrium effective potential. 
Then, as long as the quench-time $\tau_Q$ is finite -as in any 
practical application-
a second order phase transition takes place out of thermal equilibrium.

To modify the thermal activation picture, Zurek's proposed  that 
the relevant length scale of the pattern of
topological defects was given by the correlation length at $t^*$.
Using the scaling laws in (\ref{scaling}) this implies a quench-time dependence
of $\xi(t^*)$ as
\be
\xi(t^*) \sim \tau_Q^{\nu/(1+\mu)}\,,
\label{zu}
\ee
to be compared with eq. (\ref{ta}).

A recent generation of solid-state experiments, 
on phase transitions
in liquid crystals and in $^3He$ and  $^4He$, have ruled out the $\tau_Q$ 
independent scaling law (\ref{ta}) and are compatible with Zurek's eq. 
(\ref{zu}). The question is then how much of these results can be extrapolated
to the relativisitc high temperature environment of the early universe.

\section{Critical slowing down in the early universe ?}
In order to extend Zurek's picture to the early universe, two main facts have
to be taken into account; the expansion of the universe, and the need to
use a relativistic quantum field theory (QFT), since temperatures are 
typically much
larger than the masses of the particles. 

In a radiation dominated universe the quench time is given by 
$
\tau_Q = -(\dot{T}/T)^{-1} =2 \tau_H\,,
$
where $\tau_H$ the inverse of the Hubble parameter, 
which can be taken as a measure of the age of the universe.

$\tau_H$ is the third relevant time-scale of the 
problem, besides the time to the phase transition, $ t_C-t =\tau_Q 
\varepsilon$, and the relaxation time $\tau_0$. If, at  $\varepsilon=1$, 
$\tau_0 < 2 \tau_H$, the picture will be similar to that discussed in 
the previous paragraph, with $\tau_0$ growing larger than $(t_C-t)$ at some
time $t^*$, leading to the scaling law of eq. (\ref{zu}). 
There is however a second possibility, namely that the lifetime
of fluctuations becomes larger than the age of the universe, which happens if
$\tau_0 (\varepsilon=1)> 2 \tau_H$. If this happens, the time $t^*$ 
is not relevant any more since what counts is the time $t_{\mathrm age}$ 
when $\tau_0=2 \tau_H$.
A different scaling law is obtained in this case, namely
\be
\xi(t_{\mathrm age}) \sim \tau_Q^{\nu/\mu}\,.
\label{age}
\ee
Which of the two scaling laws is realized depends on the epoch at which the
phase transition takes place. Typically, taking a $\lambda \Phi^4$ theory with
$\lambda = 10^{-2}$, eq. (\ref{zu}) ( eq. (\ref{age})) is realized for
$T< 10^{11} {\mathrm GeV}$ ($T> 10^{11} {\mathrm GeV}$).

The need to use relativistic QFT poses more subtle questions.
First of all, we have to identify the relaxation rate. Instead of relying
on phenomenological equations of motion like (\ref{LANG}), first principles 
equations- directly derived from the QFT- have to be employed.
They have the general form
\be
\left[\frac{\partial^2 \;}{\partial t^2} + |\vec{k}|^2 + m^2 + 
{\mathrm Re} \Pi(E_k, \vec{k}) + 2 \gamma_{\vec{k}} 
\frac{\partial\;\;}{\partial t} + \cdots \right] \Phi(t,\vec{k}) = 
\zeta(t, \vec{k}) + \cdots\,,
\ee
where $\Pi(E_k,\vec{k})$ is the self-energy, 
\beq
\gamma_{\vec{k}} =\frac{{\mathrm Im}\Pi(E_k, \vec{k})}{2 E_k}\,,
\;\;\;\;\;\;\;\;\;\;\;\;\;\;\;\;\;\;\;\;\;\;\;\;\;\;
E_k^2 =  |\vec{k}|^2 + m^2 + {\mathrm Re} \Pi(E_k, \vec{k})\,,
\label{gamma}
\eeq
and the ellipses represent terms which are non-local in time (memory terms).
Now the damped plane wave solutions have the form $\Phi \sim e^{-i(E_k t
-\vec{k}\cdot\vec{x})} e^{-\gamma_k t}$. Comparing with (\ref{LANG}) we see
that the equation of motion in this case is second order in time and, moreover,
the dissipative term is proportional to the {\it imaginary part}
of the self-energy, and not to the real part, as for eq. (\ref{LANG}). 
At first sight, it is then not obvious at all whether CSD
is realized in the relativistic QFT as well. 

Indeed, $\gamma$ has been computed in perturbation theory by Parwani 
\cite{Pa}. At two-loops in the hard-thermal-loop resummed theory one gets
\beq
\gamma_{p.t.} = 
\frac{1}{1536 \pi}\, l_{qu} \,\lambda^2 T^2\,
\label{gammacl}
\eeq
where $l_{qu}=1/m(T)$ is 
the Compton wavelength.
The above result is valid as long as $T$ is much larger than any mass scale 
of the $T=0$ theory. If at $T=0$ there is spontaneous symmetry breaking
$m(T)$ has the form $m^2(T)= -\mu^2 + \lambda T^2/24$, thus giving
a critical temperature $T_C^2 = 24 \mu^2/\lambda$. As $T\to T_C$, $m(T)\to 0$
and eq. (\ref{gammacl}) diverges. 
In other words, the resummed perturbation theory result 
is completely at odds with what expected; we have critical speeding
up instead of slowing down!
 
\section{RG computation of $\gamma$}
It is well known that (resummed) perturbation theory cannot be
trusted close to the critical point, due to the divergence of  its effective
expansion parameter, ${\it i.e.} \; \lambda T/m(T)$. 
In refs.\cite{E,DP1} it was shown that the key effect which is 
missed by perturbation theory is the dramatic thermal renormalization of 
the coupling constant, which vanishes in the critical region.
The running of the coupling constant for $T\simeq T_C$
is crucial also in turning the divergent behavior of eq. (\ref{gammacl})
into a vanishing one. 

The details of the computation of $\gamma(T)$ in the framework of the 
Thermal Renormalization 
Group (TRG) of ref. \cite{DP1} can be found in \cite{Pi}.

\begin{figure}[t]
\centerline{\epsfxsize=1.9in\epsfbox{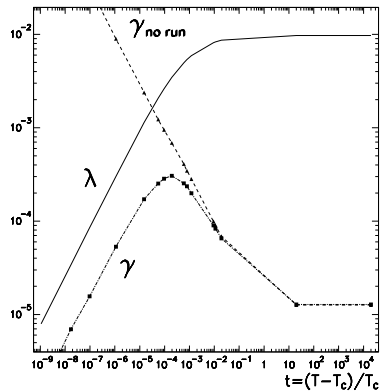}}
\caption{Temperature dependence of the coupling constant (solid line), and of 
the damping rate with the effect of the running of $\lambda$ included 
(dash-dotted) and excluded (dashed). The values for $\gamma$ have been 
multiplied by a factor of $10$.\label{fig:res}}
\end{figure}


In  Fig.  \ref{fig:res} we plot the results for the temperature-dependent 
damping rate $\gamma$ and coupling 
constant, as a function of the temperature \cite{Pi}. 
The dashed line has been 
obtained by keeping the coupling 
constant fixed to its $T=0$ value ($\lambda = 10^{-2}$), 
and reproduces the divergent behavior found
in perturbation theory (eq. (\ref{gammacl})). The crucial effect of the running
of the coupling constant is seen in the behavior of the dot-dashed line. 
For temperatures close enough 
to $T_C$, the coupling constant (solid line in Fig.2) 
is dramatically renormalized and it decreases
as
\[ \lambda(T) \sim \varepsilon^\nu \]
where $\nu \simeq 0.53$ in our approximations.
The mass also vanishes with the same 
critical index. 
The decreasing of $\lambda$ drives $\gamma$ to zero, but with a different 
scaling law \cite{Pi},
\[
\gamma(T) \sim \varepsilon^\nu \log \varepsilon\;.
\]

Taking couplings bigger than the one used in this letter ($\lambda =10^{-2}$), 
the deviation from the perturbative regime starts to be effective farther 
from $T_C$. Defining an effective temperature as
$\lambda(T)/\lambda \le 1/2$ for $T_C < T \le T_eff$ we find 
that $t_{eff}$ scales roughly 
as $t_{eff} \sim \lambda$.



\section{Conclusion} 
We have seen that Zurek's picture of second order phase transitions 
is basically valid also in the early universe, provided non-perturbative
methods are employed in order to reproduce CSD.
This problem seems to be tailored on the {\it real-time} TRG method of ref.
~\cite{DP1} better than on any
other computation method. Indeed, as we have seen, perturbation theory
fails close to $T_C$. Moreover, since we have to compute a non-static quantity
({\it i.e.} at non-zero external energy, see eq. (\ref{gamma})), neither 
lattice simulations nor  the Exact RG of the second of 
refs. \cite{E} -which are implemented in euclidean time- can be employed here.

\end{document}